\documentclass[%
 reprint,
 amsmath,amssymb,
 aps,
 prl,
floatfix
]{revtex4-2}

\usepackage[bottom]{footmisc}
\usepackage{float}
\usepackage{graphicx}% Include figure files
\usepackage{dcolumn}% Align table columns on decimal point
\usepackage{bm}% bold math
\usepackage[english]{babel}
\usepackage[unicode=true,pdfusetitle, bookmarks=false, breaklinks=false,pdfborder={0 0 0},backref=false,colorlinks=false] {hyperref}
\begin{document}

\preprint{APS/123-QED}

\title{Realignment-free cryogenic macroscopic optical cavity coupled to an optical fiber}% Force line breaks with \\

%\author{Vitaly Fedoseev$^{1*}$, Fernando Luna$^2$, Wolfgang L\"{o}ffler$^1$ and Dirk Bouwmeester$^{1,2}$}
\author{Vitaly Fedoseev$^{1,a)}$}
%\email{email:vfedoseev@physics.leidenuniv.nl}%
%\author{Fernando Luna$^2$}
%\author{Ian Hedgepeth$^2$}
\author{Matteo Fisicaro$^1$}
\author{Harmen van der Meer$^1$}
\author{Wolfgang L\"{o}ffler$^1$}
\author{Dirk Bouwmeester$^{1,2}$}
%\author{Wolfgang L\"{o}ffler and D.Bouwmeester}%
 
\affiliation{%
 $^1$Huygens-Kamerlingh Onnes Laboratorium, Leiden University, 2333 Leiden, CA, The Netherlands.\\
 $^2$Department of Physics, University of California, Santa Barbara, CA 93106, USA.
}%


%\author{Fernando Luna}
%\affiliation{
% Department of Physics, University of California, Santa Barbara, CA 93106, USA. 
%}%
%\affiliation{
%\affiliation{%


\date{\today}% It is always \today, today,

%\begin{abstract}

%Usual abstract
%\begin{description}
%\end{description}
%\end{abstract}


\begin{abstract}

We present a cryogenic setup where an optical Fabry-Perot resonator is coupled to a single-mode optical fiber with coupling efficiency above 90\% at mK temperatures without realignment during cooling down. The setup is prealigned at room temperature to compensate for the thermal contraction and change of the refractive index of the optical components during cooling down. The high coupling efficiency is achieved by keeping the setup rotation-symmetric around the optical axis. The majority of the setup components are made of Invar (FeNi36) which minimizes the thermal contraction. High coupling efficiency is essential in quantum optomechanical experiments.
\\
\\
$^{a)}$Author to whom correspondence should be addressed: vfedoseev@physics.leidenuniv.nl
\end{abstract}

\maketitle

\section{Introduction}
%To be written:
%- Why such a cavity is needed?\\

Complex optical systems with several optical elements are challenging to cool down while keeping aligned due to thermal contraction and changes in refractive index \cite{Purdy2012}. Therefore such systems are often based on optical free space access cryostats \cite{Purdy2012, Kuhn2014, Peterson2016, Rossi2018} to allow for external compensation of the optical misalignment. Alternatively, fiber coupled optical systems with cryogenic compatible actuators can be used leading to an increase in design complexity and requiring active control protocols while cooling down \cite{Verhagen2012, Fogliano2021}. Such systems are of particular interest for quantum optomechanical experiments that require low temperature environment and high collection efficiencies \cite{Aspelmeyer2014, Cohen2015, Riedinger2016, Galinskiy2020, Fedoseev2021} because the rate of thermal decoherence is proportional to the temperature of the environment \cite{Aspelmeyer2014}. In quantum optomechanical experiments detection of a scattered photon projects the mechanical system into a specific quantum state. Therefore the total collection efficiency $\eta$ of the heralding photons is essential for high fidelity operations and affects the duration of data collection $\tau$ in such experiments:  $\tau\propto 1/\eta^n$, $n\geq2$ \cite{Riedinger2018, Weaver2018a}. A constituent of the total collection efficiency is the coupling efficiency of the cryogenic optical cavity to the outside optical path, often an optical fiber. In a Helium-4 flow cryostat with free space optical access a coupling efficiency of 60\% is reported \cite{Galinskiy2020}, while for a free space optical access dilution cryostat a coupling efficiency of 85\% was achieved \cite{Peterson2017}. Keeping mode matching between a single-mode fiber and a free space cavity with cryogenic actuators is even more challenging, requiring many degrees of freedom to be controlled during cooling down. The resulting mode matching in such systems is quite limited, with reported values in the range of 30-40\% \cite{Buters2017, Xu2019}.

Here we present a fiber coupled free space 10 cm long Fabry-Perot resonator which is designed to be fully axially symmetric. This allows for room temperature prealignment of the system such that the optimal alignment is obtained at cryogenic temperatures. Without any position adjustments during a cool down 96$\pm$2\% coupling efficiency between a single-mode fiber and the optical cavity with a finesse of 12000 at the optical wavelength of 1064 nm is achieved. The elimination of actuators  does not only simplify the system but also significantly improves the mechanical stability which is of vital importance for optomechanical experiments \cite{Purdy2012}.

\section{Setup}
%- axial symmetric setup; 
%- no tip-tilt, xy shift of the optical components instead; this allows for the pre-alignment (reference about the fiber-chip)
%- 4 screws which are retracted after alignment
%ask Harmen name of the shift mechanism, is it new to him - xy translation stage
%- invar guaranties low integral thermal expansion;

Figure 1 shows the design concept of our custom-built optical setup.  The main body and all the elements in contact with the optical components are made of Invar

%Fig.1:
\begin{figure}[ht]
\centering
\includegraphics[scale=0.37]{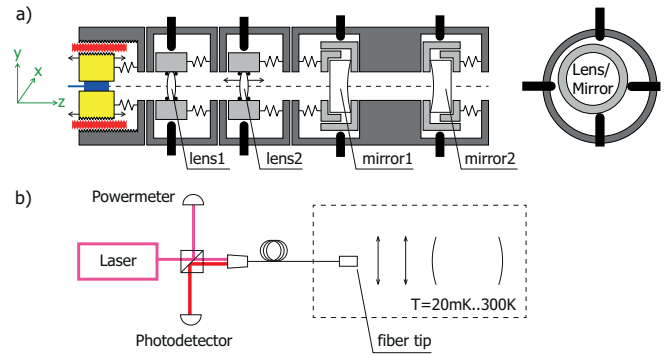}
\caption{Cryogenic setup. (a) The setup is axially symmetric and is made mostly of Invar 36. The fiber ferrule is held in a differential screw, the part shown in red can be rotated, the inner part of the screw is pushed with a leaf spring and allows for purely translational shifts of the ferrule along the optical axis. The lenses and mirrors are held in compartments (light grey) which are pushed against the setup modules (dark grey) with springs. These compartments can be shifted perpendicularly to the optical axis by 4 screws (right sketch). The screws are retracted after the alignment and the compartments are held by friction during cooling down. (b) The setup is placed in a dilution cryostat. The fiber tip of the cryogenic setup is not angled and is not antireflection-coated, the light reflected from the tip interferes with the light both reflected and leaking from the cavity.}
\end{figure}

\begin{table*}[ht]
\caption{\label{tab:table3} Positions and properties of the optical elements calculated using Gaussian beam propagation.}
\begin{ruledtabular}
\begin{tabular}{m{5cm}cccc}
    & lens1 & lens2 & mirror1 & mirror2 \\ 
 \hline
 Focal length, mm & 4.5 & 48 & 25 & 25 \\  
 Distance from the fiber tip, mm & 4.2 & 40 & 94 & 190 \\      
 Material & D-ZK3 & N-BK7 & fused silica & fused silica\\
\end{tabular}
\end{ruledtabular}
\end{table*}

%Fig.2:
\begin{figure*}[!ht]\centering
\includegraphics[scale=0.8]{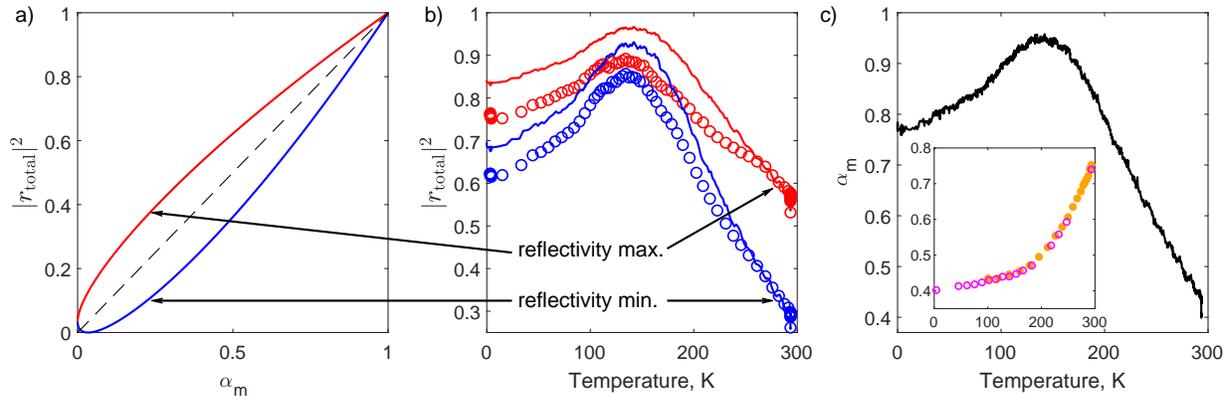}
\caption{Total reflectivity $|r_\mathrm{total}|^2$ far away from cavity resonances. (a) Theoretical interference of the light back-reflected from the fiber tip and mirror1 coupled into the fiber vs fiber coupling efficiency of the light back-reflected from mirror1 $\alpha_m$. The red (blue) line is the maximum (minimum) total reflectivity $|r_\mathrm{total}|^2$ for a given coupling efficiency. (b) Red (blue) lines/circles show the measured maximum (minimum) reflectivity due to the interference described in (a) during cooling down of the setup. The lines show the values extracted from measured visibility of the interference, while the circles show the reflectivity calculated as the ratio of light intensities measured by the photodetector and the powermeter with the ratio scaling corresponding to the room temperature measurements. (c) Coupling efficiency of back-reflected light from mirror1 $\alpha_m$ vs temperature extracted from the measured interference visibility. In the inset we demonstrate the reversibility of the coupling efficiency changes by showing the coupling efficiency $\alpha_m$ vs temperature for cooling down and warming up. The data is from another experimental run with a different optics positioning, magenta circles - cooling down, orange circles - warming up.
}
\end{figure*}

\noindent 36 to minimize thermal contraction. The light reaches the setup via an optical fiber (1060XP, numerical aperture 0.14) which is terminated with a ceramic ferrule, not angled. The ferrule is held by a differential screw (Invar 36) with a pitch difference of 0.1 mm, where the rotatable part is depicted in red. The inner part of the differential screw is connected to a leaf spring (Invar 36), which allows translation along the optical axis without rotation, and eliminates movement of the inner part due to the tolerances in the threading. Tests of the differential screw showed that the light originating from the fiber ferrule and passing through the first lens only deviates by no more than 1 mm at a distance of 1 m from the ferrule when the screw is rotated by one turn.

%\begin{table*}%[ht]
%\begin{center}
%\begin{tabular}{ m{5cm} | m{2cm} m{2cm} m{2cm} m{2cm}}
%%\begin{tabular}{ c | c c c c}
%   & lens1 & lens2 & mirror1 & mirror2 \\ 
% \hline
% focal length, mm & 4 & 48 & 25 & 25 \\  
% distance from the fiber tip, mm & 3.8 & 44 & 94 & 188 \\      
% material & fused silica & N-BK7 & fused silica & fused silica\\
%\end{tabular}
%\end{center}
%\end{table*}
Another important design feature is that the lenses and mirrors are held in compartments (light grey) which are pressed towards the body of the setup (dark grey) along the optical axis with springs. The compartments can be shifted perpendicularly to the optical axis by four screws (80 PTI). When the alignment is finished, the screws are retracted and removed which does not change the position of the compartments, and the compartments are held in place by friction. A small shift of the lenses perpendicularly to the optical axis leads to small changes of the position and direction of the light between lens2 and mirror1, effectively acting as a periscope.

The system is designed to be fully axially symmetric: the thermal contraction and change of optical properties of the elements lead to the shift of the waist position of the cavity mode relative to the fiber tip only along the mechanical axis, which can be precompensated before cooling down. The system can be aligned in such a way that the light in the setup travels along the mechanical axis of symmetry. In a conventional design with fixed mirrors and a periscope the cavity mirrors are, to some degree, randomly positioned not guaranteeing the coincidence of the optical and mechanical axes. This issue is especially important for a cavity close to concentric (as here) where the angle between the two axes in the cavity $\alpha$ is given by $\tan{\alpha}=e/(2R-L)$ in the symmetric cavity case, $e$ is the error of positioning of the mirrors, $R$ is the mirrors radius of curvature, $L$ is the distance between the mirrors. For our case of $L=94$ mm, $R=50$ mm and a typical error of $e=0.1$ mm would lead to $\alpha=0.017$ rad or 0.8 mm shift of the beam center on the mirror. 

\section{Alignment procedure}
Positions of the optical elements leading to the theoretically perfect mode matching are shown in Table I. The aligning procedure starts with positioning mirror1 on the mechanical axis of the setup (z-axis). Light going from the fiber through lens1 and a pinhole at the position of mirror2 on the mechanical axis is imaged with an intensity calibrated camera. This allows to position lens1 such that the optical and mechanical axes coincide. Next mirror1 is added and its position is adjusted to maximize the light intensity coupled back to the fiber. This defines the final position of mirror1.

Next, lens2 is placed at the calculated optimal position (see Table I). The fiber ferrule z-position, (x,y) positions of lens1 and lens2 are adjusted to maximize the light intensity coupled back to the fiber. The fiber tip is not anti-reflection coated and is not angled, and 3.4\% of light is reflected back into the fiber. This reflection interferes with the light reflected from mirror1. The visibility of this interference allows to calculate the coupling efficiency of the light back-reflected from mirror1 [see Fig. 2(a)]. The fine adjustment of the fiber ferrule z-position and the lens positions (ferrule z-position, (x,y) positions of lens1 and lens2) are made to minimize the interference visibility.

Now, mirror2 is added. First, its position is changed to maximize the transmission of the 0,0 transverse mode, fine adjustment is achieved by minimizing the 0,1 and 1,0 transverse Hermite-Gaussian modes transmission. The difference in the beam diameters of the light coupled to the cavity and the cavity mode is best visible in the non-zero transmission through 0,2 and 2,0 transverse Hermite-Gaussian modes. This gives an indication whether lens2 is positioned properly.

Lens2 can be shifted along the z-axis and is fixed by two threaded retaining rings. Final position of lens2 is found through repeating the two previous steps: for a set of different z positions of lens2 z-position of the ferrule and (x,y) positions of lens1 and lens2 are adjusted to maximize the light reflected from mirror1 and coupled to the fiber, and mirror2 position is adjusted to minimize the 0,1 and 1,0 couplings. The lens2 z-position with minimum 0,2 and 2,0 transmission is the optimal one. In the end of this procedure we achieved a configuration where 98.5\% of light is transmitted through the 0,0 mode and the rest is almost equally distributed between modes 0,1, 1,0, 0,2 and 2,0 modes. Transmission through other higher-order modes is at least one order of magnitude lower.

Finally, the setup is precompensated for the changes due to cooling down: in the ideal axially symmetric case the change of the optical properties of the components and thermal contraction effectively result in a shift of the waist z-position of the cavity mode at the fiber and its spot size. The spot size change has negligible effect on the fiber-cavity mode overlap. This makes it possible to precompensate for the cooling down effects only by shifting the fiber ferrule along the optical axis outwards from the cavity relative to the optimal room temperature coupling configuration. (x,y) positions of the lenses are readjusted to maximize the coupling of the light reflected from mirror1 to the fiber after the ferrule shift to compensate for small beam-pointing changes mentioned above. In the experiment reported here, the ferrule was retracted by 32$\pm$3 $\mu$m before the cool down. The retraction for this cool down is an estimate of the required precompensation based on the two previous cool downs. Each cool down provides a better estimate of the required precompensation for the next cool down.

\section{Measurements}
% are anti-reflection coated and
The fiber tip is not anti-reflection coated and is not angled, therefore the reflection signal measured by the photodetector is the result of interference of the reflections from the fiber tip, mirror1 and mirror2. This allows us to accurately measure the change of the coupling efficiency of the light back-reflected from mirror1 to fiber mode $\alpha_m$, and coupling efficiency of the cavity mode to fiber mode $\alpha_c$. In this article, we define "coupling efficiency" of a field $E$ as the intensity mode overlap of the fiber field $E_f$ and $E$ diminished by the losses $\epsilon$:

\begin{equation}
\alpha=\frac{|\int E^*E_fdA|^2}{\int E^*_fE_fdA\int E^*EdA}(1-\epsilon)
\end{equation}

In particular, if the total intensity losses on the way from the fiber tip to mirror1 are $\epsilon_\mathrm{single}$ (say, a dust particle on the fiber tip and not mode-matched reflections from lens1, lens2 and back mirror1 surfaces) then

\begin{equation}
\begin{aligned}
\alpha_m&=\frac{|\int E_m^*E_fdA|^2}{\int E^*_fE_fdA\int E^*_mE_mdA}(1-\epsilon_\mathrm{single})^2\\
\alpha_c&=\frac{|\int E_c^*E_fdA|^2}{\int E^*_fE_fdA\int E_c^*E_cdA}(1-\epsilon_\mathrm{single})
\end{aligned}
\end{equation}

\begin{figure*}[!ht]
\centering
\includegraphics[scale=0.8]{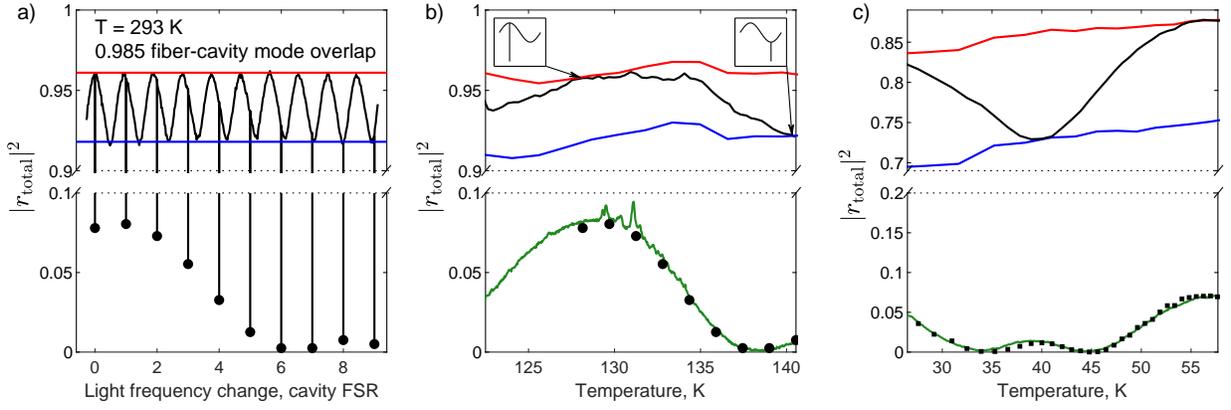}
\caption{Total reflectivity $|r_\mathrm{total}|^2$. (a) Room temperature light frequency scan in the case of optimal alignment before the precompensational shift of the ferrule. The cavity resonance dips were measured separately. (b) A particular cavity resonance is traced during cooling down (after the precompensation). The black line shows the off-resonant reflectivity in the vicinity of the traced resonance, red (blue) - maximum (minimum, excluding cavity resonances) reflectivity when the laser frequency is swept over 1 FSR of the fiber tip - mirror1 system. The insets show the relative phases between the fiber tip - mirror1 system and the cavity. Green - resonance reflectivity of the traced cavity resonance. Black circles show the resonance reflectivity values mapped from the room temperature alignment shown in a). The very similar reflection properties of the optimal alignment at room temperature and at 130 K shows that the setup alignment at 130 K is very close to the best achievable alignment at room temperature. (c) Black, red, blue and green - same as in b). Black squares represent theoretical reflectivity dip values calculated with $\alpha_c=94$\%.
}
\end{figure*}
The expression for $\alpha_m$ includes the term $(1-\epsilon_\mathrm{single})^2$ as the light traveling from the fiber and back is experiencing the losses $\epsilon_\mathrm{single}$ twice. The total reflectivity amplitude is given by

\begin{equation}
\begin{aligned}
&r_\mathrm{total}=-r_1+\frac{(1-r_1^2)Me^{i\phi_1}}{1-r_1Me^{i\phi_1}} \mathrm{,  where}\\
&M=\sqrt{\alpha_m}r_2-\frac{\alpha_c t_2^2r_3e^{i\phi_2}}{1-r_2r_3e^{i\phi_2}}
\end{aligned}
\label{eq_main}
\end{equation}

and $r_1, r_2, r_3$ are reflectivity amplitudes of the fiber tip, mirror1, mirror2, respectively. $t_2$ is transmissivity amplitude of mirror1, $\phi_1=2\pi/\lambda\times2L_\mathrm{t1}$ with $\lambda=1064$ nm and $L_\mathrm{t1}$ being fiber tip - mirror1 optical distance, $\phi_2=2\pi/\lambda\times2L_{c}$ with  $L_{c}$ being  cavity length. According to the specifications, mirror1 and mirror2 have an intensity transmissivity of 250 ppm; the scattering and absorption loss is not specified, but our measurements are consistent with 30 ppm. $r_1=\sqrt{0.034}$ based on the Fresnel coefficients for silica.

First, let's consider optical frequencies far away from the cavity resonances (detuning from the nearest cavity resonance $>$10 cavity linewidths, cavity finesse is 12000). For these frequencies the measured reflection signal is the interference of the system fiber tip - mirror1. In this case Eq. \ref{eq_main} can be simplified:

\begin{equation}
r_\mathrm{total}=\frac{-r_1+\sqrt{\alpha_m}r_2 e^{i\phi_1}}{1-r_1r_2 \sqrt{\alpha_m} e^{i\phi_1}}
\label{eq_simple}
\end{equation}

The theoretical total reflectivity $|r_\mathrm{total}|^2$ minimum and maximum values ($\phi_1=0$ and $\phi_1=\pi$ respectively) are shown in Fig. 2(a) as a function of coupling  efficiency of back-reflected light from mirror1 $\alpha_m$. For known $r_1$ and $r_2$ the visibility of this interference is a function of $\alpha_m$ only. This allows us to accurately estimate the coupling efficiency $\alpha_m$ by measuring the visibility of the interference between the fiber tip and mirror1 reflections during cooling down of the setup, no calibration is needed. The minimum and maximum of the total reflectivity can be calculated from the measured visibility data, these are plotted by the solid lines in Fig. 2(b).

The total reflectivity can also be calculated directly as a calibrated ratio of the reflection signal and the powermeter signal, see Fig. 1(b). Resulting minimum and maximum reflectivity values are shown by circles in Fig. 2(b). This estimation assumes that the transmission of the optical fiber going to the cryostat does not change with time/temperature. There is a fiber connector in the cryostat at the 4 K plate, which might explain the difference between the two estimates of the total reflectivity. The coupling efficiency of the light back-reflected from mirror1 $\alpha_m$ calculated using the visibility data is shown in Fig. 2(c). There is an initial discontinuity in $\alpha_m$ at room temperature. This change occurred during evacuation of the cryostat vacuum chamber causing a change in properties of the optical elements. $\alpha_m$ reaches 95\% at 140 K setting an upper bound on the total losses: $\epsilon_\mathrm{single}\leq0.025$. Lens1, lens2 and the back sides of the mirrors have anti-reflection coatings with a reflectivity of $\sim$0.2\% per surface (specs) setting the lower bound on the total losses: $\epsilon_\mathrm{single}\geq0.01$.

The change of the coupling efficiency with temperature is reversible: $\alpha_m$ is the same for a given temperature during cooling down and warming up as shown in the inset of Fig. 2(c) (data from previous cool down).

Next, we consider interference of the fiber tip, mirror1 and mirror2. The highest cavity mode overlap we could achieve at room temperature is 0.985 as stated in the section Alignment Procedure. A laser frequency scan covering 9 cavity resonances is shown in Fig. 3(a). A fit of Eq. \ref{eq_main} gives $\alpha_c=0.965$ for this data, providing an estimate of the losses $\epsilon_\mathrm{single}=0.02\pm0.005$ (0.985/0.965-1). The interference pattern of fiber tip - mirror1 results in the sinusoidal reflectivity variation. The optical path from the fiber tip to mirror1 is 6\% longer than the cavity length, which agrees well with the ratio of the cavity and fiber tip - mirror1 free spectral ranges of $8.5/8$ extracted from Fig. 3(a). 

During the cooling down the laser frequency was actively controlled in a way to trace the same resonance of the cavity. To achieve this, the laser frequency was scanned over 100 cavity linewidths once a second and the laser frequency was corrected to keep the cavity resonance approximately in the middle of each frequency scan. The reflectivity on cavity resonance is shown by green lines in Fig. 3(b) and 3(c). Every 10 minutes the laser was scanned over 1 fiber tip - mirror1 FSR to obtain the minimum and maximum reflectivity (blue and red lines in Fig. 3(b) and (c)).

The data plotted by black, blue and red lines in Fig. 3(a) and (b) allows to estimate $\phi_1$ from Eq. \ref{eq_main}. This phase information allows to map the reflectivity dip values from the optimal room temperature alignment onto the $130\pm10$ K temperature interval as depicted by black circles in Fig. 3(b). To do this $\phi_1$ is estimated for each resonance from Fig. 3(a) and the resonance reflectivity values from Fig. 3(a) are plotted for temperatures corresponding to the same set of $\phi_1$ values estimated from Fig. 3(b). Minimum and maximum reflectivity values together with the reflectivity on cavity resonance are practically the same for the optimal alignment at room temperature (before precompensation) and at around 130 K which can be seen by direct comparison of the data from Fig. 3(a) and (b). Therefore, the cavity indeed can be precompensated to reach near optimal alignment at cryogenic temperatures. 

During the two temperature intervals shown in Fig. 3(b) and (c) $\phi_1$ was changing by approximately $2\pi$. The data with $\phi_1$ spanning $2\pi$ is enough to extract a single value of $\alpha_c$ consistent with Eq. \ref{eq_main}: 0.96 for Fig. 3(b) and 0.94 for Fig. 3(c). This analysis was performed on 7 additional temperature intervals giving 7 additional single values $\alpha_c$, these estimates of the cavity coupling efficiency are shown by blue circles in Fig. 4.

%Fig.4:
\begin{figure}%[ht]
\centering
\includegraphics[scale=0.9]{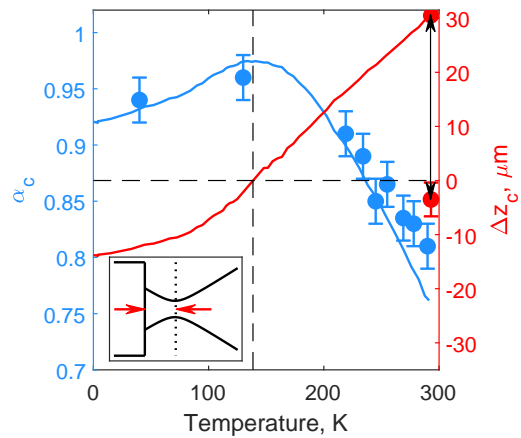}
\caption{Cavity mode coupling efficiency (left axis). Blue circles - fits to the measured reflection data. Red line - shift of the cavity mode waist at the fiber tip (see inset) calculated from the data of Fig. 2(c) under the assumption that the misalignment due to cooling down is axial symmetric. Such a shift would result in the cavity mode coupling efficiency depicted by the blue line. The arrow represents the ferrule shift made at room temperature to precompensate for the changes due to cooling down. Dashed lines are guides to the eye.
}
\end{figure}

%blue circles
%fig 2c -> red line
%red line -> blue line
%coincidence of blue line and circles
%coincidence arrow and red line

Next, we check the hypothesis whether the cavity misalignment is axisymmetric during the cool down. Under this assumption the mode overlap of the back-reflected light from mirror1 with the fiber mode is a single variable function of the distance $\Delta z_m$ between the fiber tip and the waist of the mirror1 back-reflected light, depicted by the red arrows in the inset of Fig. 4. We estimated that the beam waist radius of the back-reflected light changes negligibly small during the cool down. The back-reflected light from mirror1 mode overlap with the fiber mode can be calculated as $\alpha_m/(1-\epsilon_\mathrm{single})^2$, see Eq. 2, and then used to calculate $\Delta z_m$ as all parameters of the Gaussian beams are known.

The distance between the fiber tip and the waist of the light leaked from the cavity at the fiber tip [see inset Fig. 4] $\Delta z_c$ is twice smaller than $\Delta z_m$ because the light leaked from the cavity goes through lens1 and lens2 once, while the light back-reflected from mirror1 goes through lens1 and lens2 twice: $\Delta z_c=\Delta z_m/2$. Calculated values of $\Delta z_c(T)$ are shown in Fig. 4 by the red line. This estimate $\Delta z_c(T=293\:\mathrm{ K})=30$ $\mu m$ agrees well with the precompensation shift of the fiber ferrule of 32$\pm$3 $\mu$m. Further, calculated values of $\Delta z_c(T)$ are used to calculate $\alpha_c(T)$ shown by the blue line in Fig. 4. These values are consistent with the previous estimates of $\alpha_c(T)$ based on the reflection dip measurements (blue circles) confirming the hypothesis of mainly axisymmetric misalignment of the setup during cooling down. The calculated fiber ferrule precompensation shift to reach the highest cavity mode coupling efficiency at mK temperatures is 45 $\mu$m.

Next, we discuss vibrational properties of the setup. The setup was cooled in a dilution cryostat with cooling provided by a pulse tube cryocooler known to introduce high levels of vibrations. We used a mass-spring system \cite{DeWit2019} acting as a fourth-order mechanical low pass filter with corner frequency of 50 Hz. The filter was placed between the 20 mK plate and the cryogenic cavity setup. The optical linewidth of the cavity was 130 kHz, the Pound-Drever-Hall lock was extremely stable with visible mechanical vibrations at 2-4 kHz just above the noise floor of our detection scheme while using 20 $\mu$W of probe laser intensity. Due to the monolithic design of the cryogenic cavity the most likely part susceptible to vibrations at 2-4 kHz are the mirror spring-loaded holders.

This setup is designed for quantum optomechanical experiments \cite{Fedoseev2021} where temperatures below 1 K and high collection efficiencies are required \cite{Riedinger2016, Galinskiy2020}. For this, we have made an opening in the center of the cavity where an Invar 36 based holder can be mounted with a Si chip with a transparent high-stress SiN membrane. The membrane can be positioned at the waist of the cavity mode (beam waist radius 60 $\mu$m) and vibrational modes of the membrane are coupled to the light inside the cavity. The cavity finesse with the membrane is $\sim$10000, requiring the membrane to maintain its orientation perpendicularly to the optical axis during cooling down within $\sim10^{-4}$ rad according to our experience. Cooling down the setup with the membrane shows the behavior of the reflectivity dip similar to Fig. 3(c), indicating that the loss due to membrane tilt is below $10^{-5}$ per membrane transmission. To maximize the collection efficiency of the photons scattered on the vibrations of the membrane, mirror2 will be exchanged for a 10 ppm transmission mirror to send the light leaking from the cavity predominantly towards the fiber. We would like to note that the absence of an anti-reflection coating on the fiber tip does not diminish the collection efficiency in the case of the ideal cavity - fiber mode matching and transmissivity of mirror1 being much larger than the transmissivity of mirror2: the scattered photon will leak from the cavity preferentially through mirror1 and will be either transmitted to the fiber or reflected from the fiber tip, and after entering the cavity will leak again towards the fiber.

%the mode matching between the fiber and the light reflected from mirror1 (Fig. 2c) is along the optical axis only (right axis). 
\section{Conclusions}
We developed and characterized a single-mode fiber-coupled cryogenic optical cavity enabling cavity mode coupling efficiency above 90\% at mK temperatures. In particular, the setup does not require realignment during cooling down, due to the axisymmetric design and low expansion of the body made from Invar. The optics are aligned at room temperature to precompensate for the thermal contraction and change of refractive index during cooling down. The cavity mode coupling efficiency increases upon the start of cooling down from 81$\pm$2\%, reaches 96$\pm$2\% at 130 K and stabilizes at 92$\pm$2\% at dilution cryostat temperatures. As there are no actuators in the setup and due to its monolithic design, it has increased stiffness which reduces its susceptibility to vibrations in the cryostat. 

The setup is designed to operate in the membrane-in-the-middle configuration \cite{Thompson2008} for optomechanical experiments in the quantum regime \cite{Fedoseev2021} where high collection efficiency is essential \cite{Galinskiy2020}.

%- pre-alignment at RT 60 percent, alignment procedure (1,1 mode)
%- change of refractive index with T;
%- procedure of measuring;
%- cavity is close to confocal: high sensitivity in XY shift;
%- setup works with a membrane
%refractive in index change of the first lens has greater impact then invar 

\section{Acknowledgements}
We would like to thank K. Heeck for useful discussions of the practical design of the setup. This work was supported by the NSF Quantum Foundry through Q-AMASE-i program award number DMR-1906325 and NSF QuIC-TAQS OMA-2137740, and the NWO Gravitation-grant Quantum Software Consortium - 024.003.037.

\section{Author Declarations}
\subsection{Conflict of interest}
The authors have no conflicts to disclose.

\section{DATA AVAILABILITY}
The data that support the findings of this study are available from the corresponding author upon reasonable request.

\bibliography{library}% Produces the bibliography via BibTeX.

%\nocite{*}%

\end{document}